\documentclass[amsmath,amssymb]{revtex4}
\usepackage[parfill]{parskip}    
\usepackage{graphicx}
\usepackage{amssymb}
\usepackage{epstopdf}
\usepackage{caption}
\usepackage{color}
\usepackage{graphicx}
\usepackage{pdfpages}

\begin{document}
\title{Preliminary test of cosmological models in the scale-dependent scenario}
\author{A. Hern\'andez--Arboleda $^{a}$\footnote{a.hernandez105@uniandes.edu.co},
\'A. Rinc\'on $^{b}$\footnote{arrincon@uc.cl}, 
B. Koch $^{b}$\footnote{bkoch@fis.puc.cl},
E. Contreras $^{c}$\footnote{ejcontre@espol.edu.ec}\ \footnote{On leave from Universidad Central de Venezuela.},
 P. Bargue\~no $^{a}$\footnote{p.bargueno@uniandes.edu.co} }
\address{
${}^a$Departamento de F\'isica, Universidad de los Andes,
	Cra.1E No.18A-10, Bogot\'a, Colombia\\
${}^b$Instituto de F\'isica, Pontificia Universidad Cat\'olica de Chile, Av. Vicu\~na Mackenna 4860, Santiago, Chile\\
${}^c$Escuela Superior Polit\'ecnica del Litoral, ESPOL, Facultad de Ciencias Naturales y 
Matem\'aticas, Apartado Postal 09-01-5863 Campus Gustavo Galindo Km 30.5 V\'ia Perimetral, Guayaquil, Ecuador.}


\begin{abstract}
In the present work, we study for the first time a scale--dependent gravitational theory in a cosmological context in a matter--dominated era. In particular, starting from the Einstein Hilbert action with cosmological constant assuming scale--dependent couplings, we derive the corresponding effective Friedmann equations for the model and we solve them. We analyse in detail our results by comparing them with observational $\Lambda$CDM data as well as the very well-known  
running vacuum models. Finally, we have provided, in figures, the evolution of the Hubble parameter respect to the redshift as well as the gravitational coupling respect to the Hubble parameter and they show an agreement with the current observations.
\end{abstract}

\maketitle

\section{Introduction}

Cosmological observations which are obtained at very large distance scales,
are described with huge success by the theory of General Relativity (GR hereafter). In particular, the very well-know Lambda Cold Dark Matter model (abbreviated $\Lambda$CDM) 
allows to describe observational data with impressive precision \cite{Patrignani:2016xqp,Ade:2015xua}.
There are, however, reasons to go beyond this classical and in a sense minimal description
within GR.
First, with the increasingly improved observational techniques cosmology
is becoming more and more precise and thus sensible to possible deviations of
the standard  description within the $\Lambda$CDM model.
Second, there are theoretical reasons to believe that  deviations from 
this classical theoretical description should exist and that the leading deviation should 
come from effective scale dependence of the coupling parameters of the theory.
In general, scale dependence of couplings
is a very common effect in many field of physics, such as solid state, statistical, or high energy physics.
In the context of GR, the observed couplings are the cosmological ``constant'' 
$\Lambda_0$ and Newton's ``constant'' $G_0$.
In particular,
Dirac was the first one who considered the gravitational coupling as a time dependent quantity \cite{dirac1979large}. 
After Dirac's paper, other models appeared as candidates to explain why and how 
these coupling constants depend on space and time \cite{berman1991cosmological,bertolami1986time,bertolami1986brans}. 
Particularly interesting is the work of Bermann \cite{berman1991cosmological} in which a generalization of the 
Einstein field equations is obtained by promoting $\{G_0, \Lambda_0\}$ to $\{G(t), \Lambda(t)\}$.
Some measurements suggest that the coupling ``constants'' $G_0$ and $\Lambda_0$ could, indeed, vary in time \cite{barrow1996time,anderson2015measurements}. 
Following those ideas, many models have been made in order to incorporate time dependence 
 of the coupling constants $G_0$ and $\Lambda_0$ (for an incomplete list see, 
\cite{anderson2015measurements,fritzsch2015fundamental,fritzsch2016running,desai2016frequentist}). 
Usually the starting point is within the Einstein field equations for time dependent couplings. 
However, those procedures have the problem that most adhoc modifications of the field equations 
could lead to inconsistencies with the symmetries of the underlying theory.
Even though such models have proven to fit successfully to experimental data of both couplings \cite{Sola2017}, it is necessary to provide a self consistent treatment which give us effective equations from the variational principle incorporating running couplings.

We think that those issues of self consistency could be avoided in a 
new approach, where the coupling constants are treated as functions of a scale/field,
which has been proposed and successfully 
applied to cosmology and black holes physics 
 ~\cite{
 Reuter:2003ca,
 Reuter:2004nv,
 Koch:2010nn,
 Contreras:2013hua,
 Koch:2015nva,
 Contreras:2017eza,
 Koch:2016uso,
 Contreras:2018dhs,
Rincon:2018lyd,
Contreras:2018swc,
Rincon:2017ypd,
Rincon:2017goj,
Rincon:2017ayr,
Rincon:2018sgd}. 
This approach is inspired by the fact that the effective quantum action 
presents a scale dependent behavior, which is very well known in the literature. 
Thus, one has to take into account deviations from the classical (non scale dependent) solutions of
GR.
Similar ideas of models with running gravitational coupling have been considered in the context of grand unification theories \cite{Reeb:2009rm}. 

In the present work we deduce the effective Einstein field equations assuming scale (temporal) dependent couplings for a FLRW background. After that, we solve the corresponding effective Einstein field equations in some relevant cases and 
compare the solutions
with the previously obtained solutions reported in Ref. \cite{sola2014vacuum,Sola2017} 
and with observations in case these are available. 
Throughout the manuscript we will be interested in one of the main cosmological functions related to the scale factor of the Universe,
which is the Hubble parameter as a function of the redshift, $H(z)$. 
Even more, given the large number of both measurements and
theoretical computations on $H(z)$ (see \cite{Sola2017,Sola:2017znb} and references therein) in a matter dominated epoch, we focus our attention
only in this era.
The present work is organized as follows: 
after this introduction, we present the  $\Lambda$CDM model
and running-vacuum models (RV hereafter) in Section \ref{classic}, whereas our solutions (obtained from the variational principle) are collected in Section \ref{VP_matter}. Finally in Section \ref{Conclusion} we conclude summarizing our main results and we discuss the future of this investigation.

\section{$\Lambda$CDM and running--vacuum models}\label{classic}

In this section we briefly review both $\Lambda$CDM and some running-vacuum cosmological models, which have received
considerable attention given the concordance between them and recent measurements \cite{Sola2017}. A comparison between the predicted and observed Hubble factor $H(z)$ is presented for the cases here revised.

\subsection{$\Lambda$CDM model}

We first start by revisiting the classical $\Lambda$CDM model which corresponds to constant couplings, namely usual 
GR \cite{Weinbergbook}. Specifically, both the Newton's constant, $G_0$, and the vacuum energy density, 
$\rho_{\Lambda_0} \equiv \Lambda_0/(8 \pi G_0)$, are taken as constants. 
Besides, it assumes a perfect fluid such that $T^{\mu}_{\nu} = \text{diag}(-\rho_m, p_m, p_m, p_m)$. 
The unknowns are the scale factor $a(t)$, the matter density $\rho_m(t)$ and the pressure $p_m (t)$. In the simplest case, 
one can assume dust ($p_m(t)=0$) and one just needs to find the other two functions. 
The classical Friedmann equations 
are sufficient for determining those two unknown functions.

\subsection{Running Vacuum cases}

Regarding the running cases, the simplest model consists in working with the usual Friedmann equations
but replacing $\{G_0, \rho_{\Lambda_0}\}$ with $\{G(t), \rho_\Lambda(t)\}$. The equations are given by

\begin{align}
\label{bermanneqs1}
H^2 &= \frac{1}{3}\kappa(t)
\bigg[
\rho(t) + \frac{\Lambda(t)}{\kappa(t)}
\bigg]
,
\\
    \label{bermanneqs2}
\dot{H} + H^2 &= - \frac{1}{6}\kappa(t) 
\bigg[
\rho(t) + 3 p(t)
\bigg]
+ \frac{1}{3}\Lambda(t),
\end{align}
where, as usual, $a(t)$ is the scale factor for the FLRW metric, $\kappa(t) \equiv 8 \pi G(t)$ and the spatial curvature is taken as zero. Given the number of equations together with the number of unknowns, the system is not closed in general, as it will be shown in the following cases. In this sense, extra information is needed to account for a complete
description of the cosmological model under consideration.
\subsubsection{$G(t)= G_0$ and $\dot{\rho}_\Lambda(t) \neq 0$}\label{Sola1}

The key assumption lying at the heart of this case is promoting only $\Lambda_0 \rightarrow \Lambda(t)$ in Einstein's equations, 
namely
\begin{align}\label{case1_sola}
G_{\mu \nu} + \Lambda(t)g_{\mu \nu} &= \kappa_0 T_{\mu \nu}.
\end{align}
In this case, an exchange of energy between matter and vacuum takes place. The unknowns are $a(t)$, $\rho_m(t)$ and $\Lambda(t)$.
Therefore, Eqs. \eqref{bermanneqs1} and \eqref{bermanneqs2} 
allow to solve for the  unknowns 
provided  an extra input
on $\rho_{\Lambda}$ is demanded.
 A common choice is \cite{Solaproc2014}
\begin{equation}
	\label{dark energy density ansatz}
    \rho_\Lambda (t) = \frac{\Lambda(t)}{8\pi G_0} = m_0 + m_2H(t)^2,
\end{equation}
where $m_0$ and $m_2$ are coupling constants and the Hubble parameter $H(t)$ is defined according to $H(t) \equiv \dot{a}(t)/a(t)$. As pointed out recently 
\cite{Sola2017}, the above expression has been suggested in the literature from the quantum corrections of QFT in
curved spacetime. The running parameter $m_0$ is of order 
$10^{-3} M^{4}$, while $m_2 \sim 10^{-4} M^{2}$ where
$M$ is an average (large) mass scale (see, for example, 
\cite{Shapiro2003,Espana2004,Sola2015} and references therein).\\
Although the present case is considered in Refs. \cite{Sola2010,Sola2016}, in these references the authors do not present an 
explicit solution for $H(t)$, which is the observable we are interested in, as previously commented. Therefore, it is interesting to present the analytical solution corresponding to this case, which is
\begin{align}
\label{sola1}
H(t) &= \frac{2 \mathcal{M} }{8 \pi  m_2-3}
\tan \bigl(\mathcal{M} \left(2 c_1+t\right)\bigl)
-
\frac{4 \pi  m_1}{8 \pi  m_2-3},
\end{align} 
where we have defined the auxiliary function
\begin{align}
\mathcal{M} &=  \sqrt{2 \pi m_0 \bigl(8 \pi  m_2-3\bigl)- \bigl(2 \pi  m_1\bigl) ^2}.
\end{align}
As a final comment, note that the scale factor $a(t)$ as well as the remaining unknowns $\Lambda(t)$ and $\rho_{m}(t)$ can be easily obtained from the expression \eqref{sola1}.

\subsubsection{$\dot{G}(t) \neq 0$ and $\rho_\Lambda(t) = \rho_{\Lambda_{0}} $}\label{Sola2}

In this case, we will consider the vacuum energy density as a constant value. However we promote $\{G_0, \Lambda_{0}\} \to \{G(t), \Lambda(t)\}$. From the covariance of the Einstein's tensor we obtain
\begin{align}\label{case 3}
\bigl[
3 H(p+\rho) + \dot{\rho}
\bigl]
+ 
\chi (\rho + \rho_{\Lambda}) & =0.
\end{align}
where $\chi \equiv \dot{G}(t)/G(t)$.
In this case, the modified Friedmann field equations given by Eqs. \eqref{bermanneqs1} and \eqref{bermanneqs2}, 
combined with Eq. \eqref{case 3}, are not enough to solve for the four unknowns $a(t)$, $\rho_m(t)$, $\Lambda(t)$, $G(t)$. Even more, we can not use the ansatz \eqref{dark energy density ansatz} 
since $\rho_\Lambda$ is constant. Therefore we need to get extra information from some additional assumptions. 
One possibility, previously suggested in this context \cite{guberina2006dynamical} can be expressed as 
\begin{equation}
	\label{matter density ansatz}
    \rho_m(t) = m_4 a(t)^{-3+\epsilon},
\end{equation}
where $\epsilon$ signals a possible deviation from $\Lambda$CDM. This case is studied in \cite{guberina2006dynamical} without given an explicit expression for $a(t)$. However, by noting that
\begin{equation}
	G(t) = \left(\frac{m_4 a(t)^{\epsilon -3}+m_0}{m_0+m_4}\right){}^{\frac{\epsilon }{3-\epsilon }},
\end{equation}
an analytical expression for $a(t)$, and therefore for $H(z)$, can be obtained. However, given their length, the explicit expressions are not presented.

\subsubsection {$\dot{G}(t) \neq 0$ and $\dot{\rho}_\Lambda(t) \neq 0$} 

This is the most general
case within this framework. In addition to Eqs. \eqref{bermanneqs1} - \eqref{bermanneqs2}, the covariant conservation of 
$T_{\mu \nu}$ is assumed \cite{Solaproc2014} in order to have a closed system of equations. This new equation is given by
\begin{align}	\label{case 4}
\dot{\rho}_{\Lambda} + \chi (\rho + \rho_{\Lambda}) &= 0,
\end{align}
which can be solved, using some ansatz,
for instance, for $\rho_{\Lambda}(t)$ (see Eq. (\ref{dark energy density ansatz})).
Interestingly, the function $G(H)$ can be obtained analytically as \cite{Solaproc2014}
\begin{equation}
\label {GSola}
G(H)=\frac{G_{0}}{1+m_{2} \ln\left( H^2/H_{0}^2\right)},
\end{equation}
where $G_{0}$ and $H_{0}$ are the Newton's constant and the Hubble parameter at a given time (usually taken as the present
time), respectively.

\subsection{Discussion}
As commented in the introduction, our aim here is to make review of the RV models by comparing its predictions with experimental data for $H(z)$. But beyond that, it is important to point out that the RV cases have been studied in a much more detailed way, that is, they have been tested far beyond the concordance between the predicted Hubble factor and experimental measurements. Even more, its parameters have been restricted using very specific tests and several other experimental data (see for example \cite{sola2014vacuum} and reference therein). Nevertheless, here we will only focus on the behaviour of $H(z)$ since it its a simple way to find out whether a model beyond GR makes a consistent description of our universe or not.

\begin{figure}[h!]
        \centering
        \includegraphics[width=0.85\textwidth,height=0.35\textheight]{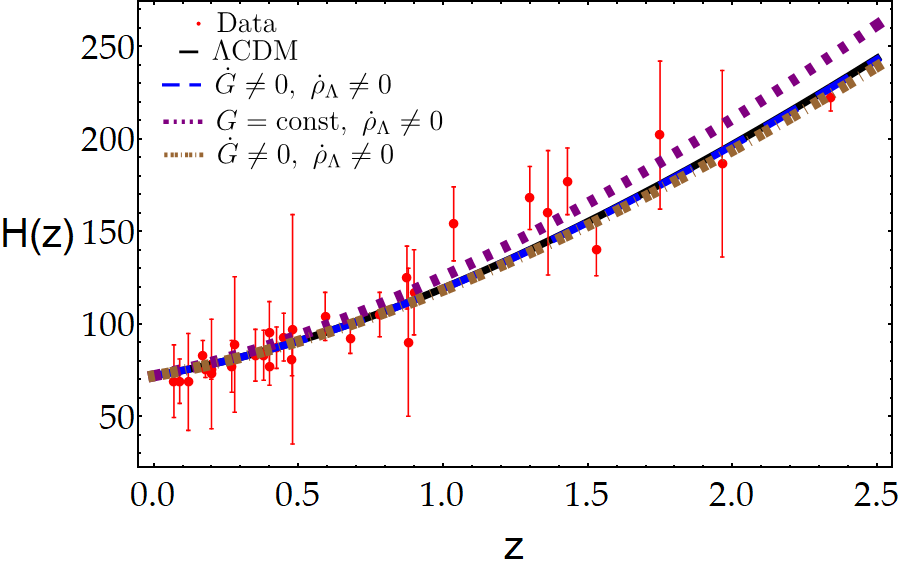}
        \caption{\label{fig1} Behaviour of $H(z)$ for the running vacuum cases considered in the present section. See text
for details.}
\end{figure}

In Fig. (\ref{fig1}) we show the comparison between experimental data (see text for details) 
and the computations under the $\Lambda$CDM and the running--vacuum model previously discussed for $H(z)$. 
The values of the parameters $m_0,\ m_1,\ m_2$ were fixed as in \cite{Sola2016}.
Given the experimental uncertainty, all the considered models reproduce fairly well the observations. Therefore, $H(z)$ does not
provide an stringent test and other observables are needed. At this point it is worth mentioning the recent work in  \cite{Sola2017}, in which the authors constrain possible running--vacuum models using the cosmological
observables SN Ia  + BAO + $H(z)$ + LSS + BBN + CMB. However, as we will show in the following section(s), $H(z)$ is not enough
to discriminate between $\Lambda$CDM, running--vacuum and variational running--vacuum models.
As we have previously commented in the introduction, in the next section we will 
deduce the Einstein field equations assuming scale (temporal)
dependent couplings for a FLRW background and we
do a preliminary test of this variational running vacuum model (VRV hereafter)
models by using $H(z)$ as the fundamental observable. In order to do this, in the next section we will present the essentials of scale dependent gravity.

\section{Scale--dependent couplings and scale setting}\label{VP_matter}
In this section we will introduce the general scale--dependent framework. 
The notation follows closely  Ref. \cite{Koch:2014joa} as well as Refs. 
\cite{Koch:2016uso,Rincon:2017ypd,Rincon:2017goj,Rincon:2017ayr,Contreras:2017eza}.
The classical couplings are the Newton and the cosmological constant, both of them being promoted to scale--dependent couplings, 
namely, $\{G_0, \Lambda_0\} \rightarrow \{G_k, \Lambda_k\}$ which, in general, depend on 
some coarse graining scale, such as
the renormalization scale, $k$. In addition, there are 
two independent  fields, which are the metric $g_{\mu \nu}(x)$ and the scale field $k(x)$. 
The effective action 
is given by
\begin{equation}\label{actk}
\Gamma[g_{\mu \nu},k] = \int \mathrm{d}^4 x \sqrt{-g} \Bigg[ \frac{1}{2 \kappa_k} \Bigl(R-2\Lambda_k \Bigl) \ + \ \mathcal{L}_{M} \Bigg].
\end{equation}  
In a top down approach one has to calculate the running couplings from
an underlying model of quantum gravity such as it is for example done in \cite{Niedermaier:2006wt,Reuter:2007rv,Percacci:2007sz,Litim:2008tt,Reuter:2012id}.
First, implications of such a scale dependence have been studied in \cite{Bonanno:2000ep,Bonanno:2006eu,Reuter:2006rg,Reuter:2010xb,Falls:2012nd,Cai:2010zh,Becker:2012js,Becker:2012jx,
Koch:2013owa,Koch:2013rwa,Ward:2006vw,Burschil:2009va,Falls:2010he,Hindmarsh:2011hx,Koch:2014cqa,Bonanno:2016dyv,Bonanno:2018gck,Bonanno:2017gji}.
If one takes the concept of scale dependence seriously, physically feasible solutions
of the effective quantum action (\ref{actk}) are obtained after choosing the arbitrary renormalization
scale $k$ as a function of the physical variables describing the system e.g. $k=k(x)$.
In particular for the case of cosmology this means that the scale $k$ should
vary with time $k=k(t)$.
Taking this into account while varying  (\ref{actk}) with respect to $\delta g_{\mu \nu}$
one obtains the effective Einstein's field equations  \cite{Reuter:2003ca,Koch:2010nn,Domazet:2012tw,Koch:2014joa,Contreras:2016mdt}
\begin{equation}
\label{generalized einstein equations}
G_{\mu \nu} + \Lambda_k g_{\mu \nu} = \kappa_k T_{\mu \nu}^{\text{effec}},
\end{equation}
where the effective energy--momentum tensor is defined according to
\begin{align}
\kappa_k T_{\mu \nu}^{\text{effec}} &\equiv \kappa_k T_{\mu \nu} - \Delta t_{\mu \nu}
\end{align}
and
where $\Delta t_{\mu\nu}$ is a tensor which takes into account the scale--dependence of the gravitational coupling. 
This term is given by the following expression
\begin{align} \label{deltatmunu}
\Delta t_{\mu\nu} &= G_k \Bigl(g_{\mu \nu} \square - \nabla_{\mu} \nabla_{\nu}
\Bigl)G_k^{-1}.
\end{align}
However, for a full and self-consistent variational treatment of (\ref{actk})
one needs to include also variations of the action with respect to $k=k(x)$
\begin{align}\label{scale_eq}
\frac{\delta \Gamma}{\delta k}=0.
\end{align}
It is straight forward to show that only
the combination of both,  Eq. \eqref{scale_eq} and Eq. \eqref{generalized einstein equations},  guarantees the conservation of the stress--energy tensor \cite{Koch:2010nn,Koch:2014joa}.  
The use of equation \eqref{scale_eq} is typically quite involved such that an analytical treatment
becomes seemingly impossible.
For highly symmetric systems in the Einstein Hilbert truncation
there is, however, a nice alternative to this procedure.
For example in cosmology, symmetry dictates that the scale is only a function of time ($k=k(t)$).
Since further the couplings are only functions of the scale ($G=G_k$, $\Lambda=\Lambda_k$)
one can conclude that the couplings are only functions of time ($G=G(t)$, $\Lambda=\Lambda(t)$).
Thus, one can avoid the source of non-analyticity in (\ref{scale_eq}) by trying to solve
only equation \eqref{generalized einstein equations} directly for $a(t), \; G(t)$, and $\Lambda(t)$
and replacing relation \eqref{scale_eq} by a well motivated ansatz 
(for example for $\rho_{\Lambda}(H)$).
Doing this, one can find explicit cosmological solutions, gain analytical understanding,
and maintain the general covariance of the system.

Finally, it is worth mentioning that, in general, a solution for a scale--dependent 
model should recover the classical limit when certain running parameter is turned off 
\cite{Koch:2016uso,Rincon:2017ypd,Rincon:2017goj,Rincon:2017ayr}. However, as it will be shown later, this is not the case for certain VRV cases because the running parameter, which controls the strength of the scale dependence, can not be identified due to the 
numerical nature of the solutions (see Sects. \ref{VRV1} and \ref{VRV2} for details).

\subsection{The model}
Assuming again a flat universe (i.e. $K = 0$) and considering dust matter ($p_m (t) = 0$), 
the generalized Friedmann equations coming from \eqref{generalized einstein equations} are

\begin{align}
      \label{einstein1}
H^2 \bigg[ 1- \frac{\chi}{H}\bigg] &= \frac{1}{3}\kappa(t)\Bigl[  \rho + \rho_{\Lambda}\Bigl],
\\
      \label{einstein2}
\dot{H} + H^2\bigg[\frac{3}{2} - \frac{ \chi}{H}\bigg]& - \frac{1}{2}\Bigl[\dot{\chi} - \chi^2 \Bigl] = \frac{1}{2}\kappa(t)\rho_{\Lambda}.
\end{align}
The previous generalized Friedmann equations are the fundamental equations to be solved in all the cases considered in this section. Nevertheless, some of the cases that we will deal with will have more than two unknowns and therefore will require additional equations in order to solve for all the unknowns. In all such cases, we will either assume the covariant conservation of the energy momentum tensor $T_{\mu \nu}$ or a perturbation ansatz for the baryonic energy density $\rho_{m}(t)$. We will justify the assumption of such additional information whenever we use it.
In the rest of the work we will explore some particular cases of these VRV models, trying to constrain
their validity by computing $H(z)$ and comparing both with previous RV models and experimental data, as previously discussed. When
possible, analytical solutions for $H(t)$ will be given. For an appropriate comparison with RV models, the same cases are considered.

\subsection{$G(t)=G_0$ and $\dot{\rho}_\Lambda(t) \neq 0$} 
If $G(t) = G_0$, then $\Delta t_{\mu\nu}=0$ and the VRV model is identical to the RV model previously discussed in subsection (\ref{Sola1}).

\subsection{$\dot{G}(t) \neq 0$ and $\rho_\Lambda = \rho_{\Lambda_{0}}$ }\label{VRV1}

In this case, we assume the same ansatz for the baryonic energy density as we used in Section \ref{Sola2}, that is
\begin{equation}
    \rho_m(t) = m_4 a(t)^{-3+\epsilon}
\end{equation}
in order to have enough information to solve the system. We use this ansatz in order to follow as close as possible the protocol usually implemented in RV models to solve the equations in each one of the cases. Even more, since we are introducing a parameter $\epsilon$, we interpret it as a perturbation parameter with respect to the $\Lambda$CDM model, therefore, its numerical value will be small.
Taking into account the above, the system of differential equations to solve consist of just the generalized Friedmann equations, and in this particular case they are given by
\begin{align}
\label{1}
a^2 \left(3 H (H-\chi)- m_0 \kappa(t)\right)=  m_4 \kappa(t) a^{\epsilon -1},
\\
\label{2}
 m_0 \kappa(t)-2 \dot{H}+2 H \chi-3 H^2+\dot{\chi}-\chi^2=0,
\end{align}
After an appropriate manipulation of equations \eqref{1}, \eqref{2} we get 
\begin{align}\label{mamalon}
G(t) = \frac{3 H^2}{8 \pi  \left(\frac{m_4 a^{\epsilon -3} \left(\dot{H}-(\epsilon -2) H^2\right)}{\dot{H}+2 H^2}+m_0\right)}.
\end{align}
It is worth mentioning that an analytical solution for the scale factor $a(t)$ was not obtained due to the complexity of the equation.
\subsection{ $\dot{G}(t) \neq 0$ and $\dot{\rho}_\Lambda(t) \neq 0$.}
\label{VRV2}
This is the most general case within the VRV framework. In addition to Friedmann equations, we assume the covariant conservations of the energy momentum tensor $T_{\mu \nu}$ in order to have enough equations to solve for the unknowns $a(t),\ G(t)$ and $\rho_{m}(t)$. We make this assumption since in this general case, we need to have the usual conservation of energy-momentum in order to properly compare the results of this case to the $\Lambda$CDM model.
Therefore, the corresponding equations in this case are
\begin{align}
	\label{111}
H^2 \bigg[ 1 - \frac{\chi}{H}\bigg] &= \frac{1}{3}\kappa(t) \bigl( \rho + \rho_{\Lambda}\bigl)
\\
	\label{222}
\dot{H} + H^2\bigg[\frac{3}{2} - \frac{ \chi}{H}\bigg]& - \frac{1}{2}\Bigl[\dot{\chi} - \chi^2 \Bigl] = \frac{1}{2}\kappa(t) \rho_{\Lambda},
\\
3 H \rho &+\dot{\rho} =0
\end{align}
where, again, the first two equations correspond to the generalized Friedmann equations and the last one is the usual conservation of the energy momentum tensor. 
After manipulating the previous equations, introducing the dependences $G(H(t)),\ \Lambda(H(t))$, considering the ansatz for $\rho_{\Lambda}(t)$ given in Eq. \eqref{dark energy density ansatz}, and 
doing the change of variables $t \rightarrow z$, we can solve numerically for the function $H(z)$.

\subsection{Discussion}
\begin{figure}
        \centering
        \includegraphics[width=0.85\textwidth,height=0.35\textheight]{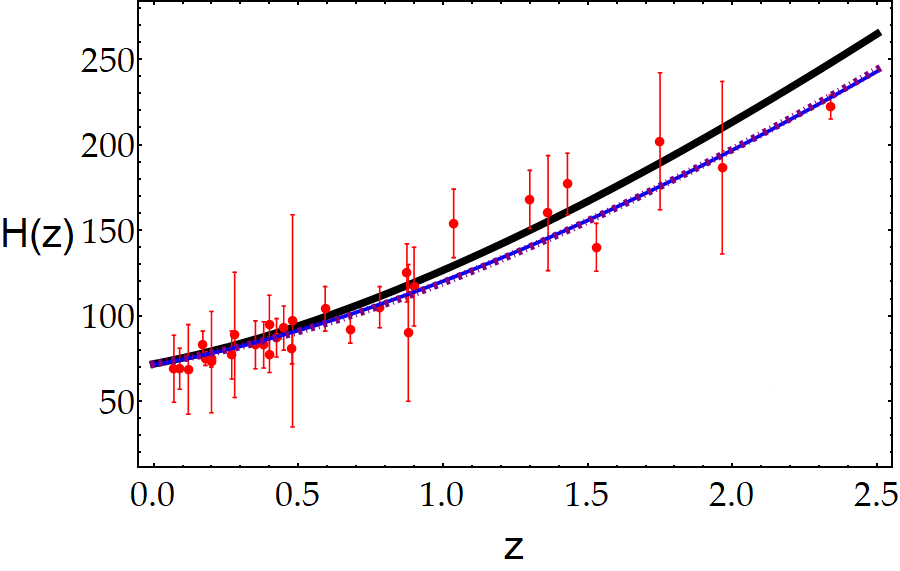}
        \caption{\label{fig2} Behaviour of $H(z)$ for the cases $\dot{G} \neq 0$ in the VRV model. The case $\dot{G} \neq 0, \rho_{\Lambda} = \text{cte.}$ is shown in blue while the case $\dot{G} \neq 0, \dot{\rho_{\Lambda}} \neq 0$ is shown in purple. Both cases show a very good agreement with the experimental data.}
\end{figure}

\begin{figure}[h!]
	\centering
	\includegraphics[width=0.85\textwidth,height=0.35\textheight]{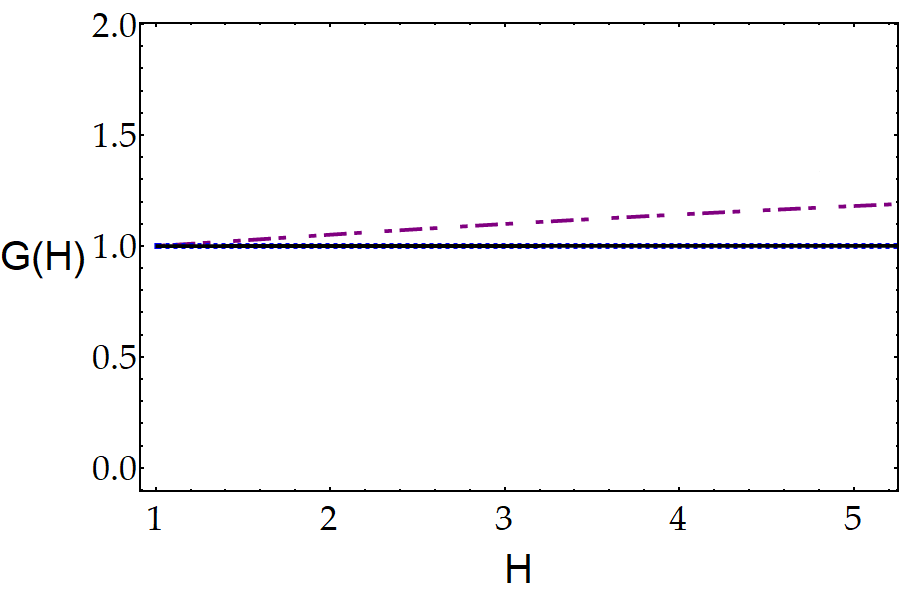}
	\caption{\label{fig3} Behaviour of $G$ in terms of the Hubble parameter $H$ for particular running vacuum and the VRV	models. The case $\dot{G} \neq 0, \rho_{\Lambda} = \text{cte.}$ is shown in blue while the case $\dot{G} \neq 0, \dot{\rho_{\Lambda}} \neq 0$ is shown in purple. The full 
	RV case is shown in black.}
\end{figure}

In Fig. (\ref{fig2}) we show the comparison between the results of cases of subsections \ref{VRV1} and \ref{VRV2} 
with experimental data, showing that both of them
fit well with the measurements. 
We note that this concordance with the expected behaviour could be due to a very slow decay
of the Newton's coupling in the context of a scale--dependent gravity, as shown in Fig. (\ref{fig3}).
To be more precise, the contribution of the Newton's coupling $G(t)$ encoded into $\Delta t_{\mu \nu}$ is the way in which the energy-momentum tensor includes any possible deviation from the classical $\Lambda$CDM model. Thus, given our solutions, the contribution of $\Delta t_{\mu \nu}$ (see Eq. \eqref{deltatmunu}) is expected to be small or at least the effect of a scale--dependent gravitational coupling on the classical solution is not dominant according to Fig. \eqref{fig3}. Regarding the Fig. \eqref{fig3} we see that the RV case is essentially constant over this scale. We think that is the reason behind the nice fit between the RV predictions, the VRV predictions and experimental data. Nevertheless, a detailed study of this hypothesis is required in order to determine its validity.

\begin{figure}[h!]
	\centering
	\includegraphics[width=0.85\textwidth,height=0.35\textheight]{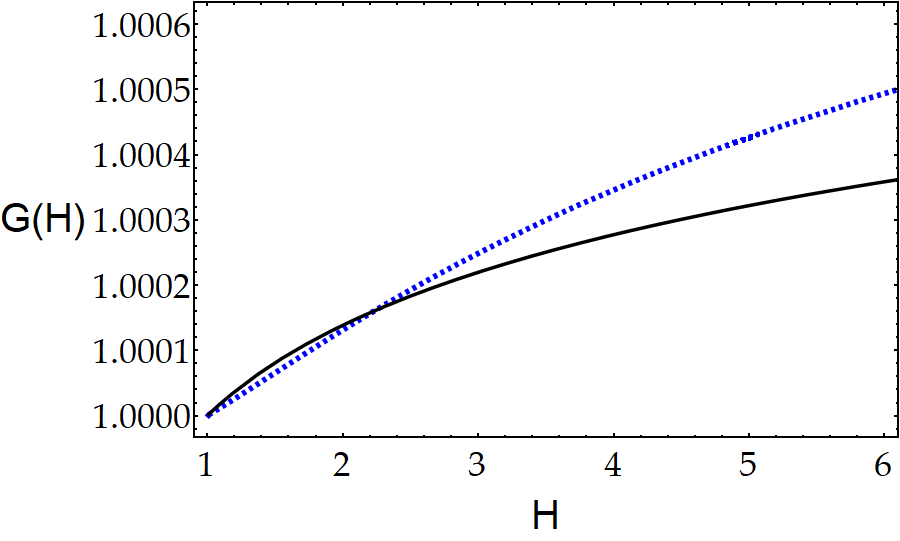}
	\caption{\label{fig4} Behavior of $G$ in terms of the Hubble parameter $H$ for the full running cases of the RV and VRV	models. The VRV case $\dot{G} \neq 0, \dot{\rho}_{\Lambda} \neq 0$ is shown in blue while the corresponding RV case is shown in black.}
\end{figure}

A more detailed comparison of the full running cases of both the RV model and the VRV model is shown in Fig. (\ref{fig4}).

Let us remark that the VRV model here considered pass the preliminary test. We think that the main reason for this is the slow decay of $G(t)$ with respect to the constant value of $G_0 = 1$. 
In particular, deviations from the classical solution are related with the rate of change $\chi \equiv \dot{G}(t)/G(t)$. Thus, in order to obtain a good agreement with the observational data, $\chi$ and $\dot{\chi}$ should be small.
 
\section{Concluding remarks}\label{Conclusion}

In this work we have introduced a scale--dependent cosmological model in a matter--dominated era. In the first part
we have 
reviewed the so--called running vacuum models, incorporating some points not previously
considered in the literature. In all these cases the model fits well with experimental data for $H(z)$ by considering
an standard equation of state for dark energy $p_{\Lambda}=-\rho_{\Lambda}$. Despite of that, it could be
more satisfactory, for a theoretical point of view, to have a running vacuum model coming from a variational principle.
In this spirit, starting with an effective action based on the usual Einstein--Hilbert one, we have promoted 
$\{G_0,\rho_{\Lambda_0}\}$ to scale--dependent couplings,  $\{G_k,\rho_{\Lambda_k}\}$ to implement this scale--setting procedure 
into a cosmological context (variational running vacuum model) for a matter--dominated era in the second part.
Specifically, and for the sake of comparison with other running vacuum models, we have limited our study to the case of
time--dependent couplings.
It is worth mentioning that we have also assumed the same ansatz for $\rho_{\Lambda}(H)$ which were
considered in the original running vacuum case. Note that, since this ansatz comes from QFT considerations in a FLRW
background, we keep it as a suitable feature also for the variational running vacuum model.
However, different parametrizations for
$\rho_{\Lambda}(H)$ coming from other approaches to quantum gravity could be assumed. 

Within the variational procedure, we have studied several cases and a brief summary of the key points in each one of them is shown in Table \ref{summary}.
	\begin{center}
		\captionof{table}{Several cases of the VRV model.}\label{summary}
		\begin{tabular}{ | c | c | c | }
			\hline
			\textbf{Case}& \textbf{Key point} & \textbf{Description} \\ \hline
			\textbf{1.} & $G(t) = \text{cte.},\ \rho_\Lambda(t) = \text{cte.}$ & Classical GR, that is, $\Lambda$CDM. \\ \hline
			\textbf{2.} & $G(t) = \text{cte.},\ \dot{\rho}_\Lambda(t) \neq 0$ &  $\nabla_{\mu}T^{\mu}_{\nu} \neq 0$ and it is the same as RV case.\\ \hline
			\textbf{3.} & $\dot{G}(t) \neq 0,\ \rho_\Lambda(t) = \text{cte.}$ & $\nabla_{\mu}T^{\mu}_{\nu} \neq 0$ and we assume $\rho_m (t) \propto a(t)^{-3+\epsilon}$ . \\ \hline
			\textbf{4.} & $\dot{G}(t) \neq 0,\ \dot{\rho}_\Lambda(t) \neq 0$ & True running, in which we assume $\nabla_{\mu}T^{\mu}_{\nu} = 0$. \\ \hline
		\end{tabular}
	\end{center}
\ \\\\
First, if $\rho_{\Lambda}$ is taken as a constant value, an analytic solution for the modified Einstein's field
equations (including $G(t)$) is obtained. It is worth mentioning that this solution does not correspond to a {\it true running}
since the classical solution can not be recovered by turning off of any parameter. 
Therefore, once the running case has been considered, the classical case is automatically discarded. 
This feature is remarkable due in previous scale--dependent problems we always can recover 
the classical solution under certain choice of the so--called running parameter of the theory.
Second, a complete numerical solution has been obtained when both $\rho_{\Lambda}(t)$ and $G(t)$ 
are taken into account into the model. Although this case corresponds to a true running, due to the numerics involved, we have 
not been able to find any parameter which could control the strength of the scale--dependence for both $\rho_{\Lambda}(t)$ and $G(t)$.
Third, regarding the comparison of these two models with the experimental data for $H(z)$, there are not significantly differences with respect to the $\Lambda$CDM model.
A possible explanation could be given in terms of the behaviour of the Newton's coupling. As the classical solution implies a 
constant value, $G_{0}$, for this coupling, we should expect the decay to be slow so that its variation, with respect to the 
classical case, is not appreciable. This turn out to be exactly the case, therefore the behaviour of the Newton's coupling with respect to the constant coupling's case could be an important point to 
take into account in order to test the viability of any extension of General Relativity, regardless of the fact that $G$ is not an observable of the model.
In addition, we note that a slower decay of $G(t)$ in the corresponding running vacuum case gives 
place to an excellent agreement with experimental data.
Beyond that, it is necessary to say that what we did in this work is only a preliminary test to establish whether the scale dependent gravity theory fits well or not to the experimental data available for the time evolution of the Hubble factor, and although this preliminary test is satisfied by our variational running vacuum model, further test are required in order to decide if this model reproduces well the behaviour of our universe at all cosmological scales or not. Specifically, we can consider solar system test of our model or linear perturbations of both the metric and the energy densities.
These and other aspects of scale--dependent cosmology are currently under study and will eventually be the object for future publication.
To conclude, our model supports that the effective action of QFT in curved spacetimes can systematically incorporate quantum corrections through scale--dependent couplings. As far as we know, this is the first preliminary test that has been done about it.

\section{Acknowledgments}

The  author  A.  R.  was  supported  by  the  CONICYT-PCHA/Doctorado Nacional/2015-21151658. 
The author B. K. was supported by the Fondecyt 1161150 and 1181694. 
The author P. B. was supported by the Faculty of Science of Universidad de los Andes, Bogot\'a, Colombia. The author A. H. A. was supported by the Faculty of Science of Universidad de los Andes, Bogot\'a, Colombia with the 2017-2 Call for Financiation of Research Projects for Master Students.

\end{document}